\documentclass[12pt]{article}
\usepackage{times}
\usepackage{geometry}
\usepackage[utf8]{inputenc}
\usepackage{enumitem,amssymb}
\usepackage{ragged2e}
\newlist{thematic}{itemize}{8}
\setlist[thematic]{label=$\square$}
\usepackage{pifont}

\usepackage{newtxtext,newtxmath}
\usepackage{txfonts}
\usepackage{caption}
\usepackage[super,comma,sectionbib]{natbib}    
\usepackage{hyperref}                          
\hypersetup{colorlinks=true,linkcolor=blue}    

\usepackage{lscape}
\usepackage{longtable}
\usepackage{ulem}       
\usepackage[dvipsnames]{xcolor}
\usepackage[T1]{fontenc}
\usepackage{ae,aecompl}
\usepackage{graphicx}	
\usepackage{amsmath}	
\usepackage{amssymb}	
\usepackage{framed,color}
\definecolor{shadecolor}{rgb}{0.63, 0.79, 0.95} 

\usepackage{eso-pic,graphicx}

\usepackage{sidecap}

\newcommand{\aap}{A\&A}
\newcommand{\aj}{AJ}
\newcommand{\apj}{ApJ}
\newcommand{\apjl}{ApJ}
\newcommand{\apjs}{ApJS}
\newcommand{\araa}{ARA\&A}
\newcommand{\mnras}{MNRAS}
\newcommand{\nat}{Nature}
\newcommand{\pasp}{PASP}
\newcommand{\ssr}{Space Sci. Rev.}

\newcommand{\Msun}{$\rm \,  M_{\odot}$}
\newcommand{\Osun}{$\rm \,  O_{\odot}$}
\newcommand{\Fesun}{$\rm \,  Fe_{\odot}$}
\newcommand{\Zsun}{$\rm \,  Z_{\odot}$}

\newcommand{\Mini}{\mbox{$M_{ini}$}}
\newcommand{\Mstar}{\mbox{$M_{\ast}$}}

\newcommand{\Teff}{\mbox{$T_{\rm eff}$}}
\newcommand{\teff}{\mbox{$T_{\rm eff}$}}
\newcommand{\logg}{\mbox{log~\textsl{g}}~}

\newcommand{\Lbol}{\mbox{$L_{\rm bol}$}}

\newcommand{\Mdot}{$Mdot$}
\newcommand{\vsini}{$v {\rm sin}\, i$}

\newcommand{\ha}{ $H \, {\alpha}$}

\newcommand{\lya}{$Ly \, {\alpha}$}
\newcommand{\Htwo}{$\rm H_2$}

\begin{document}

\AddToShipoutPictureBG*{\includegraphics[width=\paperwidth,keepaspectratio]{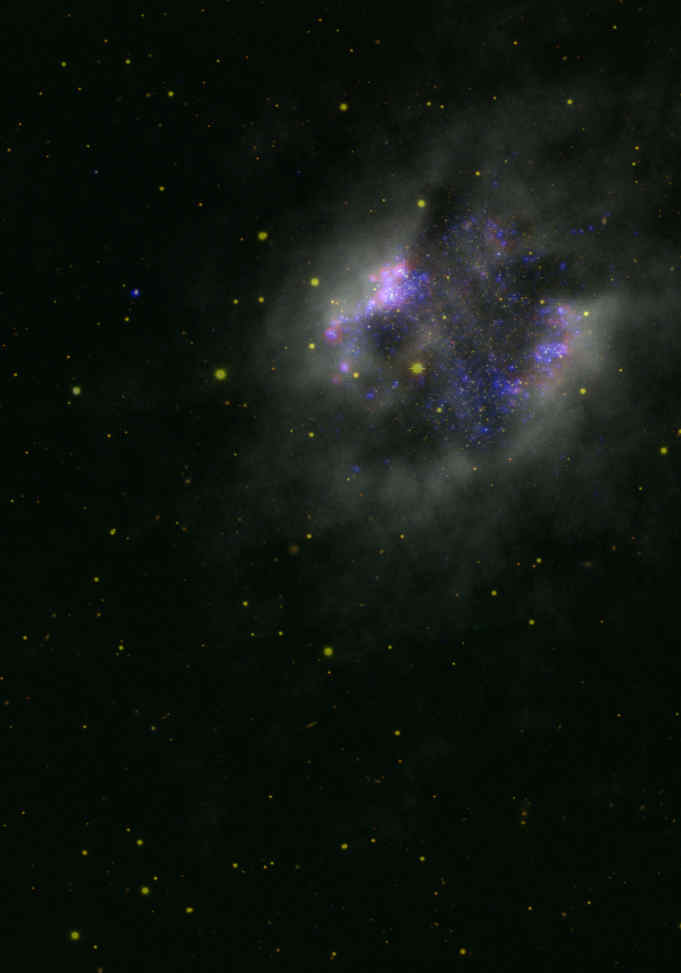}}

{ \color{White}

\raggedright
\LARGE

\Large


\vspace{1.5cm}
\hspace{-0.4cm}
\textbf{
MASSIVE STARS IN EXTREMELY METAL-POOR GALAXIES: 
A WINDOW INTO THE PAST}  \\ 

\smallskip

A white paper submitted for the Voyage 2050 long-term plan
in the ESA Science Programme
\linebreak

\vspace{5cm}
  
\normalsize

\textbf{Contact Scientist:}

Name: Miriam Garcia
\linebreak						
Institution: Centro de Astrobiolog\'{\i}a, CSIC-INTA
 \linebreak
Email: {\color{blue} mgg@cab.inta-csic.es}  \hspace*{60pt} 
Phone:  +34 91 520 2181
 \linebreak

\vspace{0.25cm}
 \justify      
 \textbf{Abstract:}
 Cosmic History has witnessed the lives and deaths of multiple generations of massive stars,
 all of them invigorating their host galaxies with ionizing photons,
  kinetic energy, fresh material
  and stellar-mass black holes.
Ubiquitous engines as they are, Astrophysics needs a good understanding
of their formation, evolution, properties and yields throughout the history of the Universe,    
and with decreasing metal content mimicking the environment at the earliest epochs.
Ultimately, a physical model that could be extrapolated to zero metallicity would
enable tackling long-standing questions such as ``What did the First, very massive stars of the Universe look like?''
or  ``What was their role in the re-ionization of the Universe?''.

Yet, most our knowledge of metal-poor massive stars is drawn from one single point in metallicity.
Massive stars in the Small Magellanic Cloud (SMC, $\sim$1/5\Zsun~) currently serve as templates for
low-metallicity objects in the early Universe, even though significant differences
with respect to massive stars with poorer metal content have been reported.

This White Paper summarizes the current knowledge on extremely (sub-SMC) metal poor massive stars,
highlighting the most outstanding open questions
and the need to supersede the SMC as standard.
A new paradigm can be built from nearby extremely metal-poor galaxies
that make a new metallicity ladder,
but massive stars in these galaxies are out of reach to current observational
facilities.
Such task would require an L-size mission, consisting of a 10m-class space telescope
operating in the optical and the ultraviolet ranges.
Alternatively, we propose that ESA unites efforts with NASA
to make the LUVOIR mission concept a reality,
thus continuing the successful partnership that made Hubble Space Telescope
one of the greatest observatories of all time.

}
\ClearShipoutPicture

\pagebreak

~~  \linebreak
~~  \linebreak

 \textbf{Proposing team:} \\
 ~~  \linebreak
\begin{tabular}{ll}
 M. Garcia & Centro de Astrobiolog\'ia, CSIC-INTA, Spain  \\ 
 \smallskip
 C. J. Evans & UKATC, Royal Observatory of Edinburgh, UK  \\  
 J. M. Bestenlehner & University of Sheffield, UK  \\
 J.C. Bouret & Aix Marseille Universit\'e, CNRS, CNES, LAM, France \\ 
 N. Castro & Leibniz-Institut f\"ur Astrophysik Potsdam, Germany  \\ 
 M. Cervi\~no & Centro de Astrobiolog\'ia, CSIC-INTA, Spain  \\ 
 A. W. Fullerton & Space Telescope Science Institute, USA  \\ 
 M. Gieles & ICC, University of Barcelona, Spain  \\
 A. Herrero & Instituto de Astrof\'isica de Canarias, Spain  \\ 
 A. de Koter & Sterrenkundig Instituut 'Anton Pannekoek', UVA, Netherlands  \\
 D. J. Lennon & Instituto de Astrof\'isica de Canarias, Spain  \\ 
 J. Th. van Loon & Lennard-Jones Laboratories, Keele University, UK  \\
 F. Martins & LUPM, Universit\'e de Montpellier, CNRS, France \\
 S. E. de Mink & Center for Astrophysics, Harvard University, USA \\   
 F. Najarro & Centro de Astrobiolog\'ia, CSIC-INTA, Spain  \\ 
 I. Negueruela & Universidad de Alicante, Spain  \\
 H. Sana & Instituut voor Sterrenkunde, KU Leuven, Belgium  \\
 S. Sim\'on-D\'{\i}az & Instituto de Astrof\'isica de Canarias, Spain  \\ 
 D. Sz\'ecsi & I. Physikalisches Institut, Universit\"at zu K\"oln, Germany  \\
 F. Tramper & IAASARS, National Observatory of Athens, Greece  \\
 J. Vink & Armagh Observatory, UK  \\
 A. Wofford & UNAM, Instituto de Astronom\'ia, Mexico  \\
\end{tabular}
\pagebreak

\justify

\section{Scientific Rationale}

\label{s:intro1}

Massive stars are stellar-size objects of broad astrophysical impact.
Born with M$>$8\Msun~ they live fast and die spectacularly, making an excellent source of fast chemical
enrichment in their host galaxy.
During a long fraction of their evolution they experience very high effective temperatures
(\Teff$\geq$20$\,$000~K, up to 200$\,$000~K in some stages) that result in an extreme ionizing
UV-radiation field.
The same UV-radiation powers supersonic winds that inject large amounts of kinetic
energy into the interstellar medium (ISM) and create ionized bubbles
and complex HII structures.
The deaths of massive stars are counted among the most disrupting events ever registered:
type Ib,c,II supernovae (SNe), pair-instability SNe,
super-luminous supernovae (SLSNe) and long $\gamma$-ray bursts (GRBs).
The surviving end products, neutron-stars and stellar-size black holes, are sites of extreme physics. 

Massive star feedback enters small and large-scale processes
spanning the age of the Universe, including the formation
of subsequent generations of stars and planets, and the chemodynamical evolution of galaxies.
Because the Cosmic chemical complexity is ever-growing after the Big Bang,
simulating these phenomena in past systems
and interpreting the available observations
demand robust models for the atmospheres, evolution and feedback of massive stars 
at ever-decreasing metallicity.
Moreover, the metal-poor ISM is more porous to UV wavelengths
and it is expected that the UV radiation of extremely
metal-poor massive stars has an impact over comparatively larger areas \citep{CAH19}.
Ultimately we need verified models for nearly metal-free very massive
stars that can be extrapolated to describe the First Stars of the Universe.

The massive stars of the Small Magellanic Cloud (SMC) constitute
the current standard of the metal-poor regime,
with a battery of observations from ground- and space-based telescopes \citep{WFC02,Mal04,ELST06,LOSC16} 
providing empirical evidence and constraints to theory.
All this is integrated into population synthesis codes
used to interpret observations of star-forming galaxies along Cosmic History.

However, the 1/5\Zsun~ metallicity of the SMC is not representative of
the extremely metal-poor regime, nor the average metallicity of the Universe past redshift $z$=1 \citep{MD14}.
The theoretical framework for lower metallicities does exist \citep{MM02,Szal15,MGB17,EIT08}
and predicts substantial differences in the evolutionary pathways 
with impact upon the time-integrated feedback and end products.
We will elaborate further on this point, but we
highlight now that one of the proposed mechanisms to reproduce
the first gravitational wave ever detected
involves the binary evolution of two metal-poor massive stars \citep{AAA16,MM16}.

This paper deals with extremely metal-poor, sub-SMC metallicity massive stars.
In the following pages we will summarize the state-of-the-art on the topic,
the exciting new scenarios that we may expect from the theoretical predictions,
how far we have reached with current observatories
and prospects for future missions in the planning.

\subsection{Formation of massive stars in metal-poor environments}
\label{s:imf}

How massive stars form remains a matter of intense investigation.
Our understanding of this topic has significant gaps
ranging from the formation of individual stars,
to how the upper initial mass function (IMF) is populated,
and whether there is a dependency on environment.
Two principal issues make the formation process markedly different from their lower-mass siblings:
the star-forming clumps must be prevented from breaking into smaller pieces,
and radiation pressure from the forming star may halt accretion.

Two main theories of star formation have emerged,
competitive accretion (radiation pressure is overcome by
  forming massive stars in the gravitational well of the whole cluster, favoring the possibility of mergers) \citep{BBC97}
and monolithic collapse (radiation is liberated via a jet) \citep{Kral09}.
At solar metallicity, they both struggle to form stars more massive than 20-40\Msun~ \citep{ZY07,HC14,TBC14}.
New promising simulations may raise this figure to $\sim$100\Msun \citep{KH18},
in better accordance with the number of known $\geq$60\Msun~ stars
in Milky Way clusters
(e.g. Westerlund~1 \citep{CNC05}, Carina \citep{S06},
Cygnus~OB2 \citep{BHC18} or the Galactic Center \citep{NdLFG17})
although still far from the $\gtrsim$150\Msun~ value registered in the
Large Magellanic Cloud \citep{Cal10} (LMC, see below).

Models also struggle to reproduce the large number of short period
binaries that are observed in
massive star populations \citep{Sal12,Sal13,KK12,BGA14}.
Fragmentation of the massive accretion disk \citep{KMY10}
is a promising way to match
the high-degree of multiplicity observed in the Milky Way \citep{Sal14,MATN19,ACNG15}
but the binaries thus formed are too wide, and an additional
hardening mechanism is required \citep{Sal17}.


The feasibility  of different scenarios of massive star formation eludes observational confrontation
since it is hard to catch forming massive stars in the act.
The most massive young stellar objects (MYSOs) have 20-30\Msun \citep{NSF18},
but at this stage a highly embedded hot core is detected where it is complicated to disentangle
the contribution of the accretion structure and the ionized gas component
with either imaging or spectroscopy \citep{SLC09}.
Few interesting links exist like \textit{IRAS 13481-6124}, an 18\Msun~ MYSO
for which VLTI-AMBER detected
a 20\Msun~ surrounding disk \citep{KHM10}.
Nonetheless, this system would only make $\sim$40\Msun~ at maximum efficiency.
The number of candidate merger events \citep{SAV16}
or merger products \citep{Vb12} is also small.

The situation should be alleviated in environments
of decreasing metallicity, since the paucity of metals would both
prevent gas from cooling and breaking up into
smaller pieces,
and 
make pre-stellar radiation-driven outflows weaker \citep{V18}.
The former argument is fundamental to support the widely-accepted concept that
the \textit{First, metal-free, stars of the Universe were very massive}.
In fact, the record-holding $\gtrsim$150\Msun~ massive stars have been found
in 30~Doradus 
at the heart of the \textit{Tarantula Nebula} in 
the LMC \citep{Cal10}, and have 0.4\Zsun~ metallicity.
Evidence of over 100\Msun~ stars has also been found in the integrated light of
unresolved, metal-poor starbursts \citep{WLCB14,SCC16}.

\begin{figure}[t]
\centering
   \includegraphics[width=\textwidth]{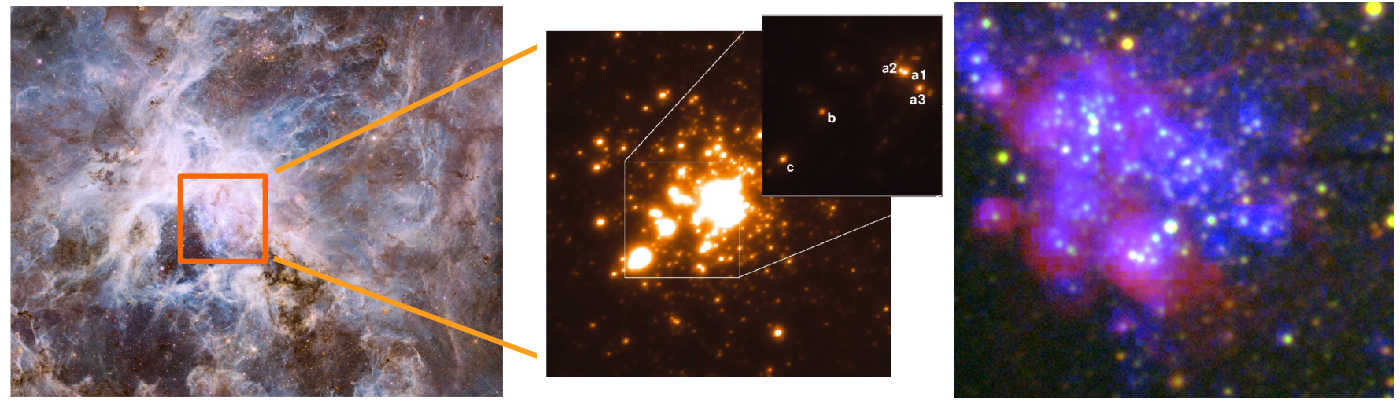}
   \caption{\footnotesize \textit{Where are the very massive stars of the Local Group?} \textbf{Left and middle panels:} The R136a cluster at the heart
     of the \textit{Tarantula nebula}, in the LMC, hosts the most massive stars known in the local Universe (\citep{Cal10}, adapted). \textbf{Right:} The Local Group 1/10 \Zsun~ galaxy Sextans~A hosts HII shells equivalent in size, but no star more massive than 60 \Msun~ has been detected.}
   \label{F:wms}
\end{figure}

\begin{figure}[t]
\centering
   \includegraphics[width=\textwidth]{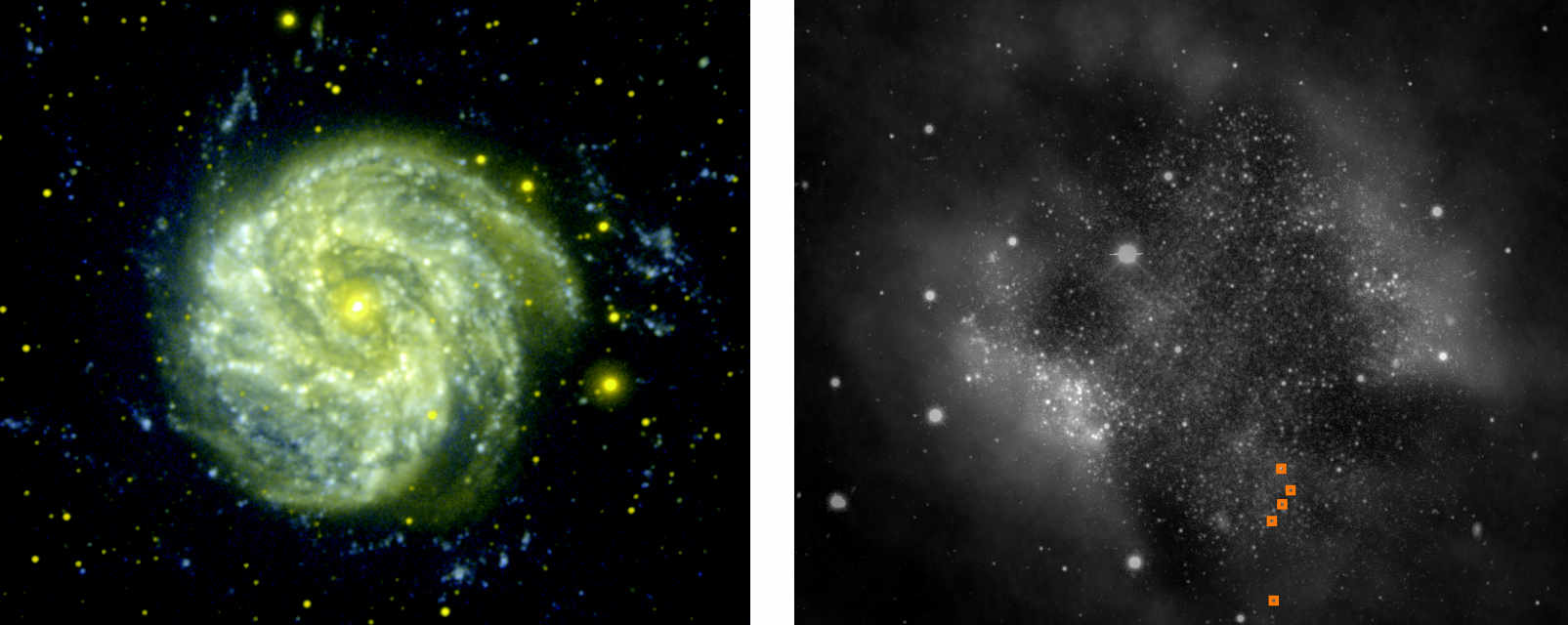}
   \caption{\footnotesize \textit{Signatures of star-formation in low gas-density environments}. \textbf{Left:} NGC~5236 and other
     extended UV-disk galaxies exhibit UV emission (hence on-going star-formation) up to 4 times beyond their optical radius. RGB composite made with FUV (blue) and NUV (green) channels  (from \citep{B18}, adapted).
     \textbf{Right:} The youngest, most massive stars ever reported in Sextans~A (squares \citep{GHN19}) are located in the outskirts of the galaxy,
     where the density of neutral hydrogen \citep{HFA12} is comparatively low.
   The HI emission is laid over a V-band image of Sextans~A.}
   \label{F:lowSB}
\end{figure}

The Local Group and nearby dwarf irregular galaxies (dIrr)
enable us
to investigate whether the upper mass limit is set to higher values as metallicity decreases
in the range of 1/7--1/30 \Osun~ (see Sect. \ref{s:ladder}).
Some of these galaxies host spectacular HII shells equivalent in size to the 30~Doradus region (Fig.~\ref{F:wms}),
but no analog to the
LMC's \textit{monster stars} has been found yet.
The most massive star reported has an initial mass of 60\Msun,
and only a handful of them have masses in the 40-60\Msun~ range \citep{Gal17}. 
It may be argued that other factors may outweigh the paucity of metals
and lead to smaller final masses,
such as the mass of the reservoir of  \Htwo, the 
local gas density or the star-formation rate \citep{GHS11},
expecting higher mass stars in denser, more active regions.
This hypothesis was also challenged by the spectroscopic
detection of OB-stars at the low stellar and gas density outskirts of Sextans~A,
two of them being the youngest and most massive stars known so far in this galaxy \citep{GHN19}.
Previous indications of young massive stars in low gas-density environments
existed, like the extended UV-disk galaxies \citep{GdP05},
but O-stars identified spectroscopically enable the unequivocal association of
UV emission at the outskirts with young massive stars (Fig.~\ref{F:lowSB}).

The density and distribution of molecular gas would be key to understand star formation
in these environments, but
this piece of the puzzle is missing.
Direct observations of cold \Htwo~ are unfeasible at 
most of the sub-SMC metallicity Local Group dIrr galaxies (0.715 - 1.3~Mpc)
and detecting CO is extremely challenging, because of the low metal content.
CO has been detected only
in a few of them \citep{ERH13,SWZ15},
overlapping the highest concentrations of stars and UV emission,
but the low-density outskirts of the galaxies have not been targeted.
The mass of molecular gas can also be estimated from 
the dust content but Herschel missed
the inconspicuous regions  at $\sim$1~Mpc (see e.g. selection by Shi et al.\citep{SAH14}).
In addition, results rely heavily on the adopted gas to dust ratios
which in turn depend on metallicity with a large scatter \citep{GLK17}.
At the moment, there is no reliable inference of the distribution
of molecular gas in the dwarf irregular galaxies of the Local Group.

A tantalizing alternative is that 
star formation could proceed directly from neutral gas.
Simulations have shown that low-density, metal-poor neutral gas can reach sufficiently low temperatures
to begin collapsing without forming \Htwo~ molecules \citep{Kr12},
breaking the link between star formation and molecular gas.
Interestingly, there is a strong spatial correlation between
HI and the location of OB stars and associations
in the dwarf irregular galaxies of the Local Group \citep{GHC10}.
Is it possible that cloud fragmentation and star formation
follows different mechanisms in dense environments hosting molecular clouds,
and sparse, neutral-gas dominated regions?
This could be a natural explanation for the occurrence of SLSNe in the outskirts of galaxies \citep{LCB14}.
If this concept was demonstrated, the simulations of the evolution of galaxies
would need to be revisited to check the significance of
the stellar mass formed
in low gas-density regions and in the outskirts of spirals.

Hence, the latest results not only highlight our poor understanding of massive star formation but also open new questions.
The joint study of resolved massive stars
and detailed maps of neutral and molecular gas
will help to unravel the relative role played by \Htwo~ and HI in star formation,
whether it changes with galactic site and metallicity, and whether it translates into different mechanisms 
that populate the IMF and the distributions of initial rotational velocity and of binaries distinctly.
Ideally, untargeted, unbiased spectroscopic censuses of resolved massive star populations would enable reconstructing
these distributions that are so important to understand star formation, establishing the local upper mass limit in particular, and checking for any dependence on metallicity and gas content.
However, observations of the required spectral quality are out of reach to current technology because:
1) main sequence O-stars at $>$ 1~Mpc are at the sensitivity limit of optical spectrographs installed at 10m ground-based telescopes;
2) the program should include near-IR observations to overcome internal extinction (significant in dIrr galaxies) \citep{GHN19} and to reach MYSOs, but both are out of reach to current IR spectrographs; and
3) the densest concentrations of gas and stars should be inspected to look for very massive stars
but these regions are hardly resolved by ground-based spectroscopic facilities even using adaptive optics.

Nonetheless if such a phenomenal database was possible
it would enable studying on-going star formation with unprecedented detail, and
re-calibrating star formation indicators once
all the stellar mass content (including extincted stars that were initially unregistered) was properly accounted for.





\subsection{Evolution, explosions and feedback}
\label{ss:evol}

The close interaction between massive stars and the Universe began with the first generation of stars.
Primordial star formation simulations and
evidence from extremely metal-poor halo stars
strongly suggest that a fraction of them were sufficiently massive and hot
as to commence the re-ionization of the Universe \citep{HHY15,Br13,FCG17}.
Ever since, signatures of their copious ionizing flux
can be seen
in highly-ionized UV emission lines
(CIV1548,1551, OIII$]$1661,1666,
 $[$CIII$]$ 1907 + C III$]$ 1909) \citep{SSV17},
indirectly in Lyman-break galaxies (LBGs),
and in a few interesting cases in the shape of \lya~emission (LAEs).
They allows us to detect galaxies
and probe the Cosmic star formation rate
out to redshift $z \sim$~10 (see e.g. introduction by \citep{WLS13}).

Understanding
massive stars with extremely low metal content
is the missing piece of information to interpret star-forming galaxies
in their many flavors, i.e. LAEs, LBGs, ULIRGs, and Blue compact dwarfs.
Their physical properties (\Teff, luminosity \Lbol, mass loss rate \Mdot) parameterized
along their evolutionary stages, will enter population synthesis
and radiative transfer codes such as
Starburst99 \citep{LSG99} and CLOUDY \citep{FKV98},
to interpret the integrated light from massive star populations \citep{WCB16}.
Armed with these tools to study the interplay between massive stars and hosts,
we can answer outstanding questions
such as 
the average ionizing photon escape fraction of galaxies, 
a crucial parameter to establish the end of the re-ionization epoch \citep{VOSH15,FPR12}.

Massive stars are born as O- or early B-dwarfs (\teff $\geq$ 20$\,$000~K),
or extreme WNh stars when very massive.
After H-burning the star undergoes a sequence of evolutionary stages that strongly
varies with the initial stellar mass.
The zoo of post-main sequence stages includes 
O and B supergiants,
red supergiants (RSGs), luminous blue variables (LBV), yellow hypergiants (YHG) and Wolf-Rayet stars (WR),
and covers an extreme temperature interval ranging from the $\sim$4$\,$000~K of RSGs \citep{DKP13}
to the $\sim$200$\,$000~K of the most extreme oxygen WRs \citep{Tral15}.
Evolutionary models must link these stages, drawing paths that depend on 
metallicity, steady/episodic mass loss (Sect.~\ref{ss:rdw}), rotational velocity and mass exchange in binary systems \citep{L12}.

Evolutionary tracks that treat rotation and mass loss
have been extensively calculated for Milky Way, LMC (1/2\Zsun), SMC (1/5\Zsun)
\citep{Brott1,Ekal12},
I~Zw18 (1/50\Zsun) \citep{Szal15,GEG19} and 
Population III stars \citep{MCK03,EMC08,YDL12}.
Significant changes are expected in the evolution of metal-poor massive stars,
some of them with tremendous impact on ionizing fluxes.
The most notable example is the incidence of chemically homogeneous evolution (CHE),
in which fresh He produced in the core
is brought to the surface by rotation-induced mixing.
A 1/5\Zsun, \Mini=25\Msun~ SMC star will usually reach the RSG stage,
but if the initial rotational velocity (\vsini) is extremely high it will evolve into
a CHE-induced
WR-like stage with \Teff$\sim$100$\,$000~K \citep{Brott1}.
This effect is magnified at lower metallicities, where
very massive stars 
can either evolve into an envelope-inflated RSG,
or stay compact in the regime of high effective temperatures
and become a Transparent Wind Ultraviolet INtense star (TWUIN) \citep{Szal15}.
TWUINs double the HI ionizing luminosity and quadruple the HeII ionizing luminosity with respect to lower \vsini~ counterparts,
and could be responsible for the extreme HeII emission detected in I~Zw18 and the $z \sim $ 6.5 galaxy CR7, currently attributed to population~III stars \citep{Keal15,SMD15}.

Aside initial mass, rotation and metallicity,
multiplicity is yet another critical ingredient
that can dramatically impact the evolution of massive stars.
Double stars with short enough orbital period ($\sim$ few years)
will exchange mass and angular momentum with their companions thus deeply altering
the future evolution of the stars involved \citep{PJH92},
end-of-life events and left-over compact products (see below).
The shortest-period systems may merge and develop extreme
properties in terms of rotation \citep{dM14}
or magnetic field \citep{SPL16}, that we are just beginning to unravel.
In systems that do not merge, the primaries (initially more massive)
will be stripped from their envelope and may become
WR stars or OB subdwarfs \citep{GMG17,GdMG18}, which are very hot
and generate significant UV excess.
The secondaries will gain mass and angular momentum,
which may trigger sheer instabilities,
enhance internal mixing and send the stars on a chemically homogeneous evolution pathway.
At the moment, we lack any information on the frequency
and period distribution of metal-poor massive binaries
beyond the Magellanic Clouds,
since only a handful of systems are known, all of them in the galaxy IC~1613
(e.g. \citep{B13}).
Nonetheless, multiplicity has a critical impact on
the global properties of massive star populations,
since it modifies the perceived mass function and upper mass limit \citep{SILM15}
and enhances/hardens the UV flux and the amount of ionizing radiation produced \citep{GMG17}.

Another fundamental aspect of the life of massive stars
is the end of their evolution.
There is a plethora of very energetic events associated to the death of massive stars:
core-collapse supernovae -SNe- (types Ib, Ic, II, IIL, IIn, IIP, IIb),
pair instability SNe, super-luminous SNe (SLSNe), electron-capture SNe, hypernovae, kilonovae and long $\gamma$-ray bursts (LGRBs).
Evolutionary models can predict the ending mechanism and leftover products
of single and binary systems \citep{WHW02,WH06,PWT17},
but observations have proven decisive to constrain and inform theory.
For instance,
pre-explosion images provided the first evidence that RSGs and LBVs can explode as SNe,
and pre-explosion spectra of the progenitor of SN1987A
showed that it was a blue supergiant, contrary to the canonical model at that time
(see e.g. \citep{Sm09,Gral13}). Likewise,
the preference of LGRBs and SLSNe for metal-poor galaxies
\citep{LCB14,CSY17}
is a clue on the specific evolution of metal-poor massive stars.
Armed with a theoretically sound, observationally constrained \textit{map of progenitors} \citep{GY07,Sm09,vD17}
where the variation with metallicity is understood,
the most energetic LGRBs and SLSNe can be used to probe the high redshift Universe,
constrain star-formation rates \citep{Peal13} or even detect the signatures of the First Stars \citep{Br13}.

The LIGO and Virgo experiments have revolutionized our view of massive
star evolution,
with already 10 in-spiraling double black hole systems
detected  during the first two observing runs \citep{AA18}.
Numbers will soon enable statistics on the distribution of black holes (BH) and neutron star (NS)
masses, that will put the predicted scenarios for the fate of massive stars to the test.
Nonetheless LIGO and Virgo have already accomplished paradigm-shifting results.
The detection of GW170817, associated with a collapsing double neutron star 
and kilonova \citep{AAA17},
linked short-GRBs with massive stars  \citep{TPE17}.
The very first gravitational wave system detected, GW150914, challenged
all we knew about the formation of black-hole systems and put
evolution of massive stars in binary systems in the spotlight.
With $36\:$M$_{\odot}$ and $29\:$M$_{\odot}$~ masses, the two BHs that merged
were significantly larger than any stellar-mass BHs known back then
($\sim$5-15\Msun) \citep{CJ14},
and those that could be formed from stellar evolution at solar metallicity
($\sim$ 20\Msun) \citep{SW14}.
This system has inspired the development of new scenarios, 
such us the CHE evolution of two metal-poor massive stars 
within their Roche Lobes avoiding mass exchange and the common-envelope phase \citep{MM16}.

The expected weak winds of metal-poor
massive stars (Sect.~\ref{ss:rdw}) provide natural means to
reach the final stages of evolution with larger stellar masses,
thus increasing the size of the ensuing BH \citep{BBF10}.
Alternatively, implementing the quenching of mass loss produced by a surface dipolar magnetic field
can also allow the star to maintain a higher mass during its evolution
and eventually form heavier stellar-mass black holes ( $>$ 25 M$_\odot$ \citep{PKM17}).
The same mechanism would enable solar-like metallicity massive stars to form 
pair-instability supernovae \citep{GME17}.
Magnetic fields play a yet to be fully characterized
role in the live of massive stars specially in metal-poor galaxies, although the relative population 
of magnetic OB-stars is relatively small \citep{WGM12}.

Constraining massive star evolution is a multi-dimensional problem.
The high incidence of massive stars in multiple systems, and the fraction that
will interact with their companions \citep{Sal12},
enriches the problem exponentially.
The way to proceed is to assemble large, multi-epoch spectroscopic datasets of
large samples of massive stars to fully cover the parameter space,
constrain their physical properties
with the most advanced stellar atmosphere models,
obtain distributions of \vsini~ and of the properties of binary systems,
and contrast against the predictions of single and binary evolutionary models.
Only a vast spectroscopic program can lead to the reconstruction of the single- and binary-evolutionary
pathways of massive stars.
Such ensembles have been built over the years in the Milky Way \citep{SD15},
and most recently in the Magellanic Clouds \citep{Eal11,COFL18b,RHO19}.
However, only a handful of massive stars have been confirmed by spectroscopy in galaxies with poorer metal-content than the Small Magellanic Cloud (see \citep{Gal17} and Sect.~\ref{s:ladder}).
At this stage no signature of CHE has been reported in these galaxies, and very few massive binaries are known.

\subsection{The winds of extremely metal-poor massive stars}
\label{ss:rdw}

\begin{SCfigure}
  \includegraphics[width=0.5\textwidth]{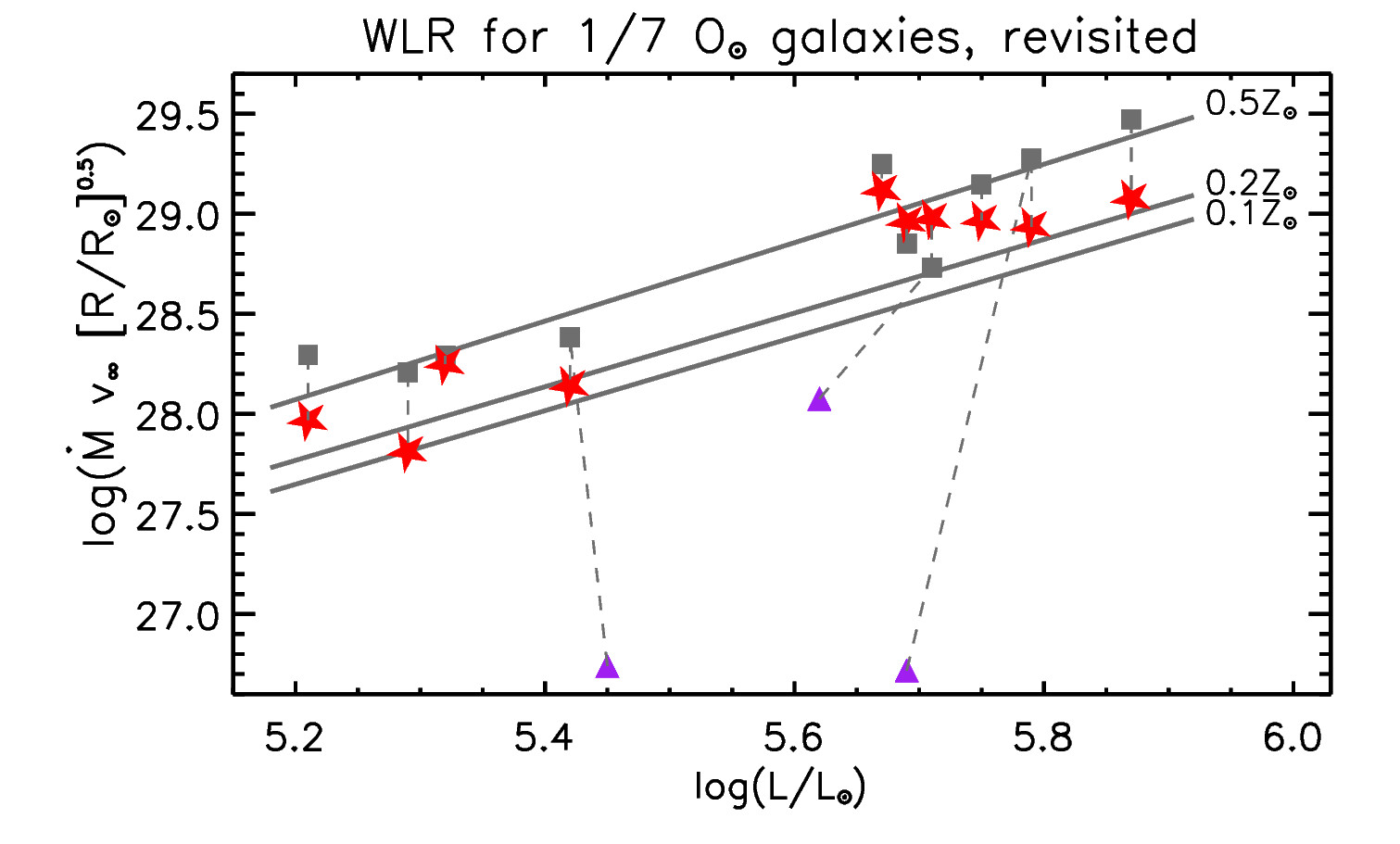}
  \caption{\footnotesize 
    \textit{The momentum carried by the wind depends on stellar luminosity and metallicity} (solid lines).
    The optical studies of  1/7 \Osun~ stars suggested that their winds were
    as strong as LMC analogs (squares). Terminal velocities from the UV revised these values downwards (stars).
    A full UV analysis resulted in wind momenta well under the theoretical prediction (triangles).
    We are far from a reliable prescription of the mass lost to RDWs by extremely metal-poor stars.}
  \label{f:wlr}
\end{SCfigure}

Stellar winds are the mechanism by which the evolution of massive stars is strongly conditioned by metallicity.
%
Massive stars experience very high effective temperatures (\teff$>$20$\,$000~K)
during a large fraction of their evolution (Sect.~\ref{ss:evol}).
In these stages the extreme UV radiation field 
exchanges energy and momentum with metal ions in the stellar atmosphere, resulting
in an outward outflow known as radiation-driven wind (RDW) \citep{LS70,CAK75}.

RDWs are particularly significant in OB-type
stars and WRs.
The ensuing removal of mass, with rates of the order of \Mdot $\sim 10^{-8} M_{\odot}/yr - 10^{-4} M_{\odot}/yr$ \citep[][]{KP00},
may be high enough to
peel off the outer stellar layers (being responsible
for the different flavors of WR-stars as successive nuclear-burning
products are exposed),
but also to alter the conditions at the stellar core and the rate of nuclear reactions.
It is because of RDWs, which inherit a strong dependence on metal content
from its driving mechanism, that two massive stars born with the same initial
mass but different metallicity can follow distinct evolutionary pathways
\citep{CM86} and result in different end-products (Sect.~\ref{ss:evol}).
RSGs, the cool evolutionary stages of massive stars, also experience outflows but the driving mechanism is different
and mass loss rates are apparently independent of metallicity \citep{GvLZ17}.

RDWs are weaker as metallicity decreases,
with models predicting
\Mdot~ $ \propto  Z^{\sim 0.85}$~ for OB-stars \citep[][]{VKL01}
and nitrogen-rich WRs \citep{VdK05}.
The theoretical relation 
has been verified
observationally down to the metallicity of the SMC \citep{Mal07}.
The winds of metal-poorer hot stars require a special formalism \citep{K02}
that should consider the shift of driving ions from Fe to CNO at Z$\leq$1/10\Zsun \citep{KK14},
implying that the wind strength may vary as processed material is brought to the surface by internal mixing.
The expectation is that at Z$<$1/100\Zsun~ winds are very weak, unless the star is very luminous,
and consequently would have very little impact on the evolution of the star.

Theory was finally confronted with observations with the arrival of multi-object spectrographs
at 8-10m telescopes.
The first efforts focused on IC~1613 (715~kpc), the closest star-forming Local Group galaxy
whose $\sim$1/7\Osun~  nebular abundance marked a significant decrease 
in present-day metallicity from the SMC.
They soon were followed by studies in the $\sim$1~Mpc away galaxies NGC3109 and WLM,
where similar nebular abundances had been measured (see \citep{Gal14} for references).
The results were unanticipated:
the finding of an LBV with strong optical P~Cygni profiles \citep{HGU10a},
an extreme oxygen WR \citep{Tal13} (WO), and the optical analysis of O-stars \citep{HGP12,Tal14}
indicated that winds were stronger than predicted by theory at that metallicity.

\begin{figure}[t]
\centering
   \includegraphics[width=\textwidth]{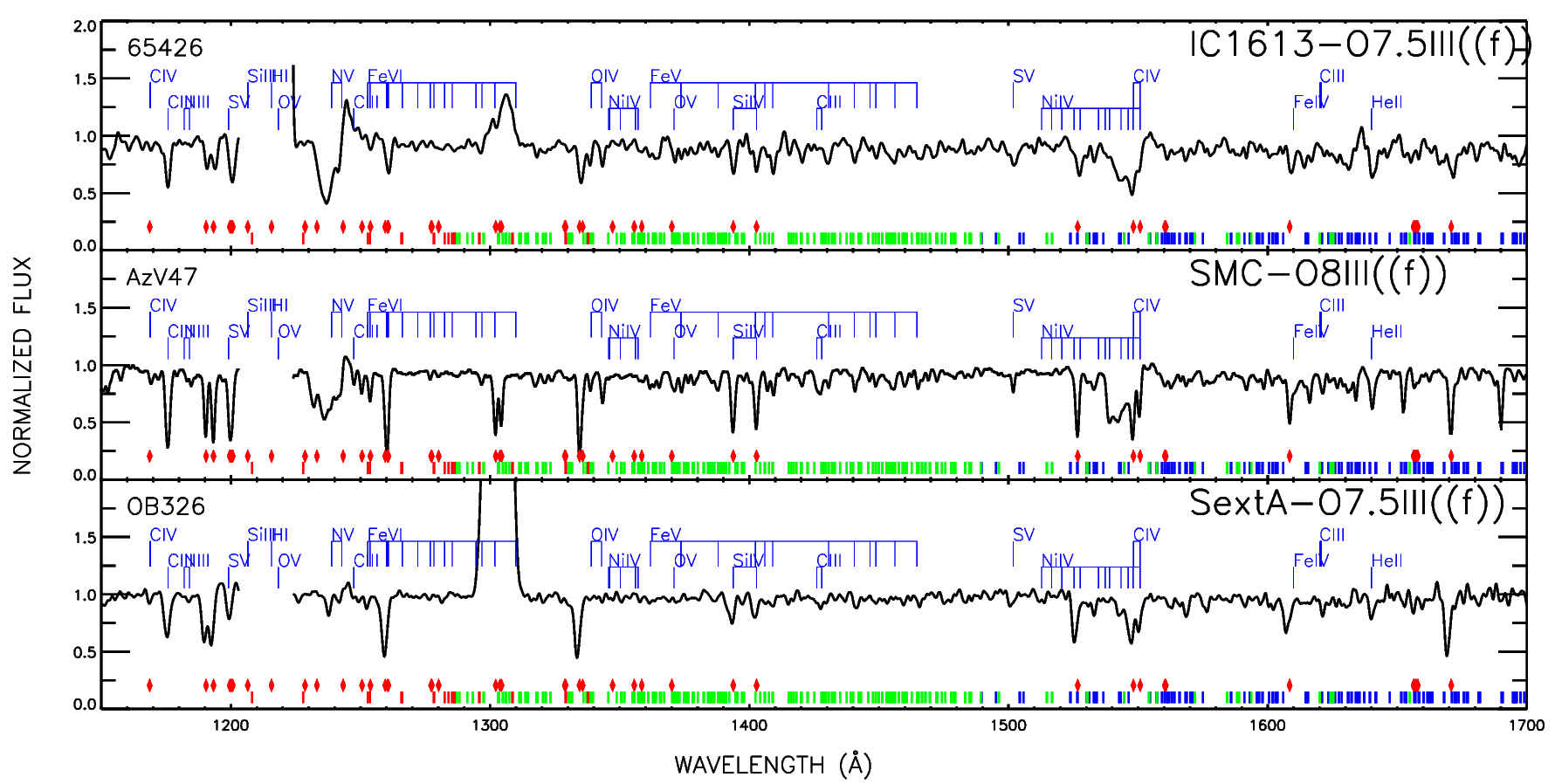}
   \caption{\footnotesize
     \textit{The UV spectral morphology reflects variations of stellar metallicity.}
     HST-COS/HST-STIS UV spectra of stars with similar spectral type (hence \Teff,\Lbol) in different Local Group galaxies.
     The pseudo-continuum at 1350--1500\AA, dominated by FeV lines (green ticks), indicates a sequence of decreasing
     Fe content from top to bottom. The wind profiles of NV and CIV decrease correspondingly.}
   \label{F:UVspec}
\end{figure}

The presence of Wolf-Rayet stars -the descendants of the most massive O-stars and
likely progenitors of Type Ib/c SNe and GRBs-
in low-metallicity galaxies is a strong indication that more mass is lost during the evolution of massive stars
than is presently accounted for.
Current single-star evolutionary models cannot explain the existence of WR stars in the SMC,
let alone the fully-stripped WO star in IC~1613.
While recent empirical results indicate only a mild dependence of
the winds of extreme WC and WO stars on metallicity  (\Mdot~ $ \propto Z^{0.25}$ \citep{TSdK16}, much weaker than nitrogen-rich WRs)
the question remains how these stars have shed their entire hydrogen envelope in previous evolutionary stages.
An interesting possibility, that brings up again the important role of multiplicity
in the life of massive stars,
is mass exchange in binary interactions \citep{GdMG18}.

UV spectroscopy by the Hubble Space Telescope (HST) played a crucial role deciphering the \textit{strong wind problem} (Fig.~\ref{f:wlr}).
The detailed analysis of UV spectral lines, more sensitive to the wind than the optical range,
yielded lower mass loss rates for O-stars \citep{Bal15}.
UV spectroscopy also showed that IC~1613's content of iron was similar or even larger than the $\sim$1/5\Fesun~ content of the SMC
\citep{Gal14} (Fig.~\ref{F:UVspec}),
superseding the 1/7\Fesun~ value scaled from oxygen.
Similar SMC-like Fe-abundances were also reported for WLM and NGC3109 \citep{Hal14,Bal15}.
While this finding
sets a reminder that metallicity cannot be scaled from oxygen abundances 
since the [$\alpha$/Fe] ratio reflects the chemical evolution of the host galaxy,
it also alleviates the discrepancy since the expected mass loss rate is larger
at the updated iron content \citep{VKL01}.

New efforts are being directed to the Sextans~A galaxy that has nebular abundances as low as
1/10-1/15\Osun~ \citep{Kal05} and similarly low stellar 1/10\Fesun~ abundances \citep{KVal04,Gal17} (see Fig.~\ref{F:UVspec}).
The first spectroscopic surveys 
have reported 16 OB stars \citep{Cal16},
but being located 1.3~Mpc away
only a handful can be observed in the UV \citep{Gal17,ullyses}.
%
Two other extremely metal-poor star-forming galaxies with resolved stellar populations have been
surveyed for O-stars, both with positive results: SagDIG (1/20\Zsun, 1.1~Mpc) \citep{G18} and Leo~P  (1/30\Zsun, 1.6~Mpc) \citep{ECG19}.
However,  the combination of distance and foreground extinction severely hampers optical observations
of O-stars in these galaxies and UV spectroscopy is basically unfeasible.
The overall sample size is insufficient and
the sub-SMC metallicity regime of RDWs remains largely unexplored.

The uncertain metallicity dependence of RDWs adds to a hotly debated question:
what is the total mass lost throughout the stellar lifetime,
and what is the  main driving mechanism?
Besides RDWs, pulsation- and rotation-driven outflows, evolution and/or mass exchange in binary systems,
and eruptions such as those experienced by Eta~Car \citep{S14}
may lead to considerable amounts of mass loss.
In fact, the concept that super-Eddington stars such as Eta Car may experience
continuum-driven winds, provides an interesting metallicity-independent mass loss mechanism \citep{vMOS}.  
These processes are very poorly understood compared to RDW even at solar metallicity, let alone among sub-SMC stars.
\textit{At the moment we simply lack any evidence to assess what is the dominant mass loss mechanism ruling the life of
extremely metal-poor massive stars}.


\section{A metallicity ladder to look back in time}
\label{s:ladder}

\begin{figure}[t]
\centering
   \includegraphics[width=\textwidth]{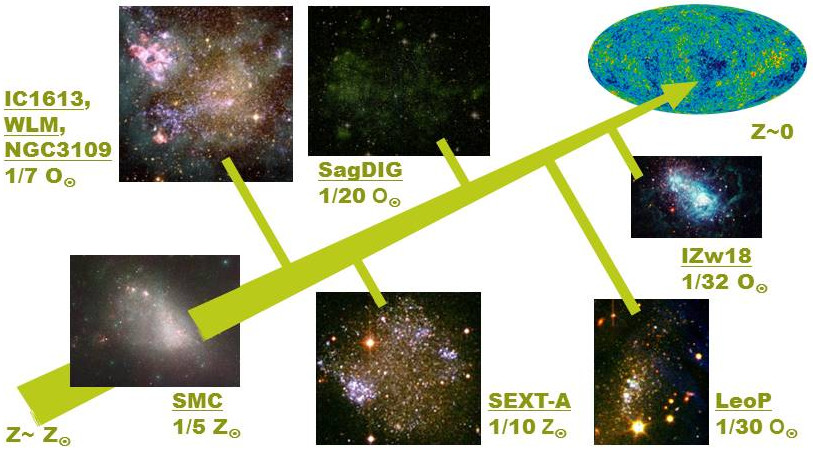}
   \caption{\footnotesize  \textit{Road-map to the early Universe}. Selected Local Group and nearby star-forming dwarf
     galaxies provide a ladder of decreasing metallicity that will allow us to study the
     physics of extremely metal-poor massive stars, and ultimately to extrapolate the properties
     of the First, metal-free massive stars.}
   \label{F:road}
\end{figure}

Massive stars are ubiquitous throughout Cosmic history
ever since the First, roughly metal-free, very massive stars.
Their ionizing and kinetic energy production is critical to many
astrophysical processes that can be counted
back to the on-set of the re-ionization epoch.
Each generation inherits the chemical composition of the cloud where it forms,
implying the existence of extremely metal-poor massive stars
in the infant Universe,
but also in pristine galaxies
where star formation was only activated recently,
or that lost their metals to galactic outflows.
Understanding the physical properties
of massive stars with extremely low metal content
is therefore crucial
to realistically compute feedback
in a significant number of environments spread through the history of the Universe.

%

\newpage

\begin{shaded}

  ~~ \newline
  \smallskip
  \textbf{Specifically, the great questions that need answering are:}

  $\bullet$~ Are the physics and evolution of extremely metal-poor massive stars substantially different from Solar metallicity analogs?
If so, what is the impact in terms of ionizing flux, yields and feedback?

$\bullet$~ Can these models be extrapolated to infer the physical properties of the First Stars (initial \Mstar, \teff, \Lbol~ and \Mdot)?
 Is it possible to detect their end-of-life events?

$\bullet$~ Does the distribution of stellar initial masses depend on metallicity?  Can extremely massive stars be expected
at the infant Universe?

$\bullet$~ What kind of death-events can be expected
from extremely metal-poor massive stars, and can any of them be detected up
to very high redshifts?

$\bullet$~ What are the evolutionary channels that lead to binary stellar mass
black holes and gravitational wave sources?
\newline
\end{shaded}


The answer to these questions relies on exceptional-quality optical and UV
spectra of a representative
sample of massive stars with sub-SMC metallicity.
Armed with the tools for quantitative spectroscopy teams around the world have been perfecting for
decades now,
accurate stellar properties
(\teff, \Mstar,  \Lbol, and wind properties) can be derived.
These will allow us to draw the evolutionary pathways of massive stars,
study the IMF in metal-poor environments
and provide more realistic previsions of feedback.
Fortunately for this quest, metallicity increases monotonically with time but not isotropically,
and some systems exist that are metal-poorer than the average present-day chemical composition of the Universe.
Unfortunately, the SMC is currently both a metallicity and distance frontier,
and a sizable leap down in metallicity requires reaching
distances of at least 1~Mpc (outer Local Group and surroundings).

Very promising galaxies with  1/10\Zsun~ (Sextans~A, 1.3~Mpc away) \citep{Cal16},
1/20\Zsun~ (SagDIG, 1.1~Mpc) \citep{G18} and 1/30\Zsun~ (Leo~P, 1.6~Mpc) \citep{ECG19}
are subject to close scrutiny with VLT, and the 10m telescopes Keck and GTC
\citep{Bal07,Eal07,Cal16,ECG19,GHN19}.
They form a sequence of decreasing metal content that will enable
understanding and parameterizing the properties of low-metallicity massive stars (Fig.~\ref{F:road}).
A crucial -yet ambitious- landmark is the 1/32 \Zsun~ blue compact dwarf I~Zw18 (18.2~Mpc) \citep{VIP98,Aal07}.
In this galaxy, very massive (300\Msun) or alternatively metal-free 150\Msun~ stars
have been suggested as possible ionizing sources producing
the extraordinarily strong observed HeII4686 nebular line \citep{Keal15}.
I~Zw18  thus represents the best chance to reach primordial-like massive stars,
and will enable studies of massive star populations in very metal-poor extreme starbursts
(very enlighting when compared with  all our compiled knowledge on 30~Doradus \citep{Eal11}).

However, the world's largest ground-based telescopes
only reach the brightest, un-reddened massive stars in
$\sim$1~Mpc galaxies after long integration times,
and even for these spectral quality is sometimes too poor as to yield
stellar parameters from quantitative analysis.
Spatial resolution is also an issue: breaking down
the population of I~Zw18 at optical wavelengths is beyond the capabilities of
even the future European Extremely Large Telescope (ELT).
The observations of stellar winds are yet more handicapped
since the intrinsic strong UV emission is dulled by extinction,
and a strong sensitivity limit is set by the relatively small mirror size
of the only observatory offering UV spectroscopy, HST.
The result is a biased  and sorely incomplete view of sub-SMC massive stars.
\textit{The reality is that we have hit the limit of current observational facilities.}


\section{Technical proposal}

The questions raised by Sects.~\ref{s:intro1} and \ref{s:ladder}
can be distilled into the following specific points:


$\bullet$~ Is the IMF universal? What is the upper mass limit?
Does it increase with decreasing metallicity?

$\bullet$~ What kind of outflows do extremely metal-poor massive stars experience?

$\bullet$~ How do their physical parameters (\teff, \Lbol~ and \Mdot) vary along evolution? What is the frequency of CHE?

$\bullet$~ What is the frequency and period-distribution of binary stars in extremely metal-poor environments?
Do they have a significant impact on feedback? Can they populate the mass distribution of double BHs and NSs
inferred from gravitational wave events?


The following subsections discuss the technical capabilities needed to
tackle these points, and a brief discussion of the
instrumentation and telescope needed to meet them.

\subsection{Technological needs for a break-through}
\label{ss:specs}

Answering the questions stated in this White Paper requires
observations of a large sample of massive stars
in sub-SMC metallicity galaxies 
in all their flavors (OB-type, WR, LBV, YHG and RSG),
with sufficient good quality as to enable detailed and precise spectroscopic analyses.
This section focuses on the OB-stars, the most challenging
to observe with current facilities.
The technical specifications set by OB-stars
will also enable observations of WRs and LBVs.
The VLT and the upcoming ELT
warrant good
prospects for YHGs and RSGs \citep{BBM15,PED17,DKL17}.

Homogeneous studies of high-quality optical and UV datasets,
such as the IACOB \citep{SD15} and VFTS \citep{Eal11} Europe-led projects,
have provided invaluable insight into
blue massive stars of the Milky Way and the Magellanic Clouds (see also \citep{COFL18b,RHO19}).
Notably, the \textit{ULLYSES} program is devoting a significant number of HST orbits
to ensure proper UV spectroscopic coverage of the SMC \citep{ullyses}.
These and other on-going efforts are consolidating our knowledge
of massive stars at the present-day, and
lay the groundwork for the kind of in-depth studies
needed to provide quantitative results on extremely metal-poor massive stars. 
\smallskip

\begin{figure}[h!]
\centering
   \vspace*{-1.1cm}\hspace*{-1.8cm}\includegraphics[width=1.15\textwidth]{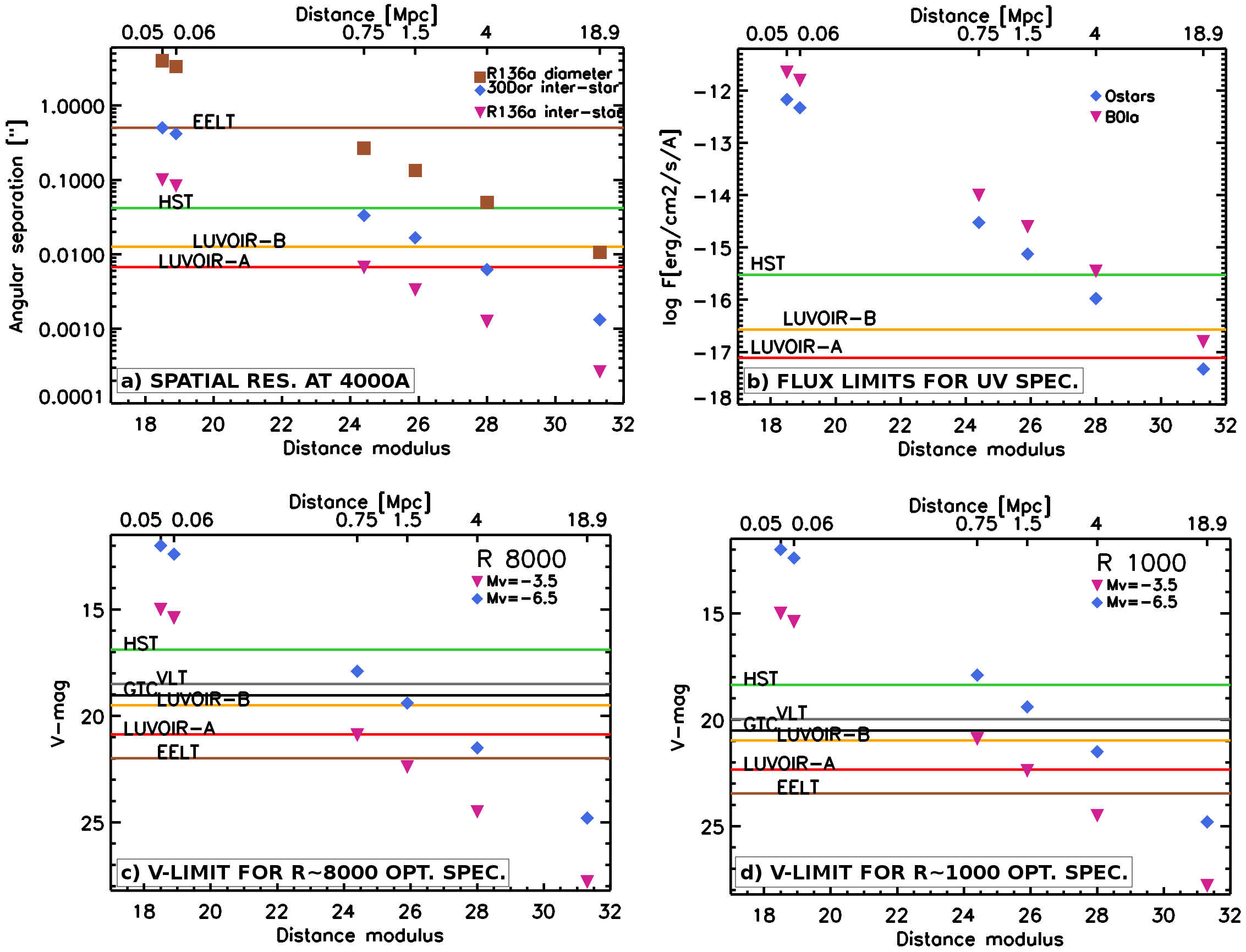}
   \caption{\footnotesize \textit{The potential of current and future instrumental facilities
       to study OB-type stars
     at landmark galaxies:
     LMC ($\sim$ 0.05~Mpc), SMC ($\sim$ 0.06~Mpc), IC~1613 ($\sim$ 0.75~Mpc), Sextans~A/Leo~P/SagDIG ($\lesssim$ 1.5~Mpc),
     the Sculptor filament and Centaurus group ($\lesssim$ 4~Mpc), and I~Zw18 (18.9~Mpc)}. \newline
     \textbf{Top-left (a): power to resolve tight stellar populations.}
     In the first column the squares mark the diameter of R136a, the compact
     cluster at the core of 30~Doradus that hosts $\sim$ 150\Msun~ stars (4$''$).
     The rhombus and the triangle mark typical 30~Doradus and inner R136a inter-star distances (0.5$''$~ and 0.1$''$~ repectively).
     The figure then illustrates the angular separation of similar structures
     at farther galaxies.
     The horizontal lines mark the diffraction limit of space facilities at 4$\,$000\AA.
     The expected performance of ELT at blue-optical wavelengths is also included as reference. \newline
     \textbf{Top-right (b): Flux limits for UV spectroscopy with R$\sim$2$\,$000.}
     Expected UV fluxes of
     O-stars (rhombuses) and B-supergiants (triangles)
     if stars were hosted by different galaxies.
     These numbers have been scaled from IC~1613 observations
     where O~stars and B~supergiants registered very similar fluxes ($\rm 3 \cdot 10^{-15}$~ and
     $\rm 1 \cdot 10^{-14}$~$erg \, cm^{-2} \, s^{-1} \,$\AA~ at 1500\AA~ respectively)
     \citep{Gal14} reflecting
     the trade-off between spectral sub-types and extinction at the time of target selection.
     The horizontal line marks the limiting flux that can be observed with HST-COS-G140L (R$\sim$2$\,$000)
     in 6 orbits
     with sufficient SNR as to enable analysis (SNR$\geq$20, $\rm 3 \cdot 10^{-16}$~$erg \, cm^{-2} \, s^{-1} \,$\AA).
     Flux limits for both LUVOIR architectures were estimated 
     scaling this value by mirror size, assuming no throughput improvement. \newline
     \textbf{Bottom: V-magnitude limits for optical spectroscopy with
       R$\sim$8$\,$000 (c, left) and R$\sim$1$\,$000 (d, right).}
     The rhombuses and the triangles provide the V-magnitudes enclosing
     O-stars hosted by different galaxies ($M_V \in \left[-3.5,-6.5 \right]$ \citep{Wal14}).
     The horizontal lines represent the magnitude reached by different facilities in
     12~hour observing time, scaled by the mirror size only, and assuming
     no throughput improvement.
     The reference for the R$\sim$8$\,$000 panel are the VLT-FLAMES observations of O~stars
     in 30~Doradus \citep{Eal11}, and GTC observations of Sextans~A \citep{GHN19}
     for the R$\sim$1$\,$000 panel. 
     An extra magnitude has been added to space facilities to simulate
     the lack of atmospheric absorption and extremely low sky-brightness.
     }
   \label{F:specs}
\end{figure}

The engine for analysis is ready, yet the observations are unfeasible with present-day instrumentation.
The key technical enabling requirements are:

$\bullet$~ \textbf{Spatial resolution of the order of 0.01$''$ at UV and blue-optical wavelengths.}
This value can resolve stellar populations out to the distance of I~Zw18 (18.9~Mpc),
disentangle 30~Doradus-like clusters throughout the Local Group ($\leq$ 1.5~Mpc)
and break-up 30~Doradus inner core, R136a, out to 750~kpc (Fig.~\ref{F:specs} top-left).
Coupled with follow-up spectroscopy, this will provide unprecedented
constraints on the IMF of dense clusters and starbursts.

$\bullet$~ \textbf{Large collecting power} to increase
sensitivity in the whole UV-optical-IR range,
enabling spectroscopy of far massive stars 
and close but extincted objects.
Optimal limiting values for different set-ups (see below) are:
V$\sim$21 for optical R$\sim$8$\,$000 spectroscopy of O-stars in the Local Group,
V$\sim$25 for optical R$\sim$1$\,$000 spectroscopy in I~Zw18,
and $\rm F_{1500A} $=$1 \cdot 10^{-17}$~$erg \, cm^{-2} \, s^{-1} \,$\AA~ for UV spectroscopy of O-stars in this galaxy
(Fig.~\ref{F:specs}).
Accessing moderately reddened OB-stars 
enables studying the processes of star formation in metal-poor galaxies
and its relation with gas density.
Strong synergies with the SPICA mission \citep{RSA18} are foreseen.

$\bullet$~ \textbf{Medium resolution (R=$\lambda/\Delta\lambda \geq$8$\,$000)
  multi-object optical and near-IR spectroscopy} to constrain stellar parameters
of massive stars and define evolutionary channels.
This configuration would mimic
the VLT-FLAMES optical survey of 30~Doradus,
that produced the most accurate characterization of LMC massive stars to date \citep{Eal11}.
The optical range ($\sim$ 4$\,$000-5$\,$500\AA~ and \ha-region)
contains the best characterized diagnostic lines constraining \Teff, \logg
and element abundances,
while near-IR spectroscopy is reserved for the most reddened population
since OB-stars are intrinsically faint in this range.
The James Webb Space Telescope (JWST) will produce
first exploratory studies in the IR,
but both its collecting area and spectral resolution
are insufficient ($\phi$~6.5m; R$_{max,NIRSpec}$=2$\,$700 whereas at least R$\sim$4$\,$000 \citep{HLA19} is needed).
An efficient optical/near-IR multi-object spectrograph
would facilitate muti-epoch observations,
which are critical to characterize spectroscopic binaries.

$\bullet$~ \textbf{Ultraviolet spectroscopy with multi-object modes}
in order to accumulate exposure time in exchange of multiplexing,
so that distant galaxies or reddened OB stars can be targeted.
The resolving power must be R$\geq$2$\,$000
to confirm the presence of winds and to
resolve the interstellar components from wind troughs when the profiles
are weak.
Observations with higher spectral resolution will enable additional constraints on
mass loss rates and the velocity field.

UV observations alone would require a space observatory, but the sensitivity
and the spatial resolution requirements also need a 10m-class telescope in space.
This is illustrated in Fig.~\ref{F:specs}, that compares these metrics
for current and future facilities:
HST ($\phi$=2.4m diameter), the ground-based telescopes VLT ($\phi$ 8m) and GTC ($\phi$ 10.2m),
the European ELT ($\phi$ 40m) and two designs for a future mission that
will be described in Sect.\ref{ss:luvoir}: LUVOIR-A ($\phi$ 15m) and LUVOIR-B ($\phi$ 8m).

While ELT's impressive collecting power will be
crucial to follow-up specific targets in the IR,
the telescope is not suitable for
large-scale studies of extremely metal-poor massive stars in the visible.
Only HARMONI among first-light instrumentation
provides partial coverage in the optical range,
missing important diagnostic lines in the uncovered 4$\,$000-4$\,$700\AA~ interval.
Even if optical coverage is considered for second-generation instruments,
adaptive optics will struggle providing diffraction-limited observations
in the optical-blue over a $\rm \geq 1' \times 1'$ field of view.
\textit{Only a large-mirror telescope in space unites both requirements of sensitivity and
  outstanding spatial resolution in the optical range.}


\subsection{The LUVOIR observatory}
\label{ss:luvoir}


A 10m-class telescope in space operating in the UV-optical-NIR ranges qualifies
as an L-size mission,
although there is a possibility that would greatly reduce the costs.
One of the mission concepts NASA is considering for its next flagship mission
meets the size and sensitivity requirements laid-out in Sect.~\ref{ss:specs}.
We propose that ESA joins as a partner.

The Large UV/Optical/IR Surveyor \citep{LUVOIR,LUVOIR2}
is a proposed multi-wavelength, large mirror telescope
operating at L2
that truly captures the heritage of HST as a broad scope observatory.
The study team is considering two architectures with different mirror size,
LUVOIR-A ($\phi$ 15m) and LUVOIR-B ($\phi$ 8m).
Both concepts are equipped with the
LUVOIR Ultraviolet Multi Object Spectrograph (LUMOS),
designed to provide high-throughput
multi-object spectroscopy at UV wavelengths.
Multiple resolution modes will be available with resolving power ranging R=500--65$\,$000.
Multiplexing will be achieved by a grid of 6 micro-shutter arrays,
with 480 $\times$ 840 shutters each,
following the design used for JWST-NIRSpec.
The multi-object capabilities of LUMOS coupled with
on-going improvements on UV detectors,
will revolutionize the
field, by enabling the first extensive characterization of the outflows of massive stars beyond the SMC.

The selection of the most ambitious design will enable UV spectroscopy
of individual stars in I~Zw18 (Fig.~\ref{F:specs} top-right).
The current specifications allow LUMOS-A to obtain
good quality spectra of I~Zw18 O-stars in about 11.5~hours (SNR=20 @ 1500\AA, R$\sim$5$\,$000) \citep{FFW17}.
Both LUMOS-A and LUMOS-B will comfortably reach
out to few Mpc distances, opening great discovery opportunities 
in the Sculptor, Centaurus and M81 Groups.
LUMOS ensures a proper characterization of RDWs
and mass loss rates in extremely metal-poor environments.

LUVOIR-A will resolve individual stars in the sparse regions of I~Zw18
in the optical and the UV (Fig.~\ref{F:specs} top-left).
Both A and B architectures will be able to dissect 
30~Doradus-like clusters -except for the densest cluster core- out to 4~Mpc.

{\footnotesize
\begin{table*}
  \caption{
    \textit{Level-zero technical specifications for an optical spectrograph onboard LUVOIR.}
  }           
\label{T:spectral}      
\centering
{\footnotesize
\begin{tabular}{|l|r|l|}
  \hline
  \textbf{Parameter}   & \textbf{Value}   & \textbf{Justification}    \\
  \hline
  Wavelength coverage                        & 3600-7$\,$000\AA & Coverage for Balmer jump, optical diagnostic lines and \ha \\
  \hline
  Resolving power                            & 1$\,$000  & Massive stars beyond the Local Group ($\geq$4~Mpc) \\
  ($\lambda/\Delta\lambda$)                 & 8$\,$000  & Massive stars in the Local Group ($\leq$1.5~Mpc) \\
                                             & 5$\,$0000 & Other (e.g. SB2 disentangling) \\
  \hline
  Faint limit, R=1$\,$000                        & V=25  & Bright O-stars in I~Zw18\\
  ~~~~~~~~~~~~~~~ R=8$\,$000                     & V=21  & Faint O-stars in Sextans~A\\
  \hline
  Field of view                              & 3$'$ $\times$ 1.6$'$  & To match the field of view of LUMOS FUV-MOS \\
  \hline
  Observing modes                            & Single-object & \\
                                             & Multi-object  & \\
  \hline
  MOS-multiplex                              & $>$10. 50 optimal & Density of targets in Local Group dIrrs\\
  \hline
\end{tabular}
}
\end{table*}
}

The true power of LUVOIR, however, resides in the combination of
sensitivity and outstanding spatial resolution
over the extent of the field of view, regardless
the wavelength range.
ELT cannot compete with the expected performance of LUVOIR in this respect.
Coupled with follow-up spectroscopy in the optical range,
these phenomenal capabilities
will enable the definite characterization of extremely metal-poor massive stars
together with unprecedented insight on the IMF of the host galaxies.

In principle, LUVOIR-A will have the required sensitivity to obtain R$\sim$8$\,$000 optical spectra
of V$\sim$21 O-type stars at 1.5~Mpc in about 12~hours (Fig.~\ref{F:specs} bottom-left).
The analysis of such dataset can provide accurate \Teff, \logg~ and abundances.
Farther galaxies require a lower resolution R$\sim$1$\,$000 mode,
enough for first estimates of stellar parameters.
LUVOIR-A could then comfortably reach V$\sim$22.5 O-stars at 4~Mpc
in 12~hours (Fig.~\ref{F:specs} bottom-right).
Reaching I~Zw18 would mean a leap of 2.5 mags that translates into a factor 10 longer
exposure times.
Such observations are feasible, but strongly advocate for multi-object capabilities.

We note that the instruments currently studied for LUVOIR do not include a
multi-object optical spectrograph working at intermediate and high spectral resolution.
This will be fundamental to many scientific cases beyond those outlined in this White Paper,
providing spectroscopic follow-up of a broad range of sources
observed by the exceptional imaging from LUVOIR.
(The French-led POLLUX study includes coverage of visible wavelengths
but at higher resolutions, R>100k, with the option of spectropolarimetry.)
We propose that ESA fills in this niche by building an optical spectrograph,
thus becoming a full partner of the LUVOIR observatory.
The basic requirements for such instrument are summarized in Table~\ref{T:spectral}.


\subsubsection{Technology challenges}

~~\\
\vspace{-1cm}

\textbf{Packaging and deployment:}
LUVOIR will build on lessons learnt by the JWST on this technological aspect,
although its larger mirror size will be an additional challenge in terms
of rocket size and mass. Active optics and mirror alignment after deployment
are also critical technological elements.

\textbf{Devices for multi-object spectroscopy in space:}
LUMOS will use a grid of micro-shutter arrays, heritage of JWST-NIRSpec.
On-flight performances will demonstrate the maturity of this technology,
but it may be interesting to test other possibilities 
offering a better trade-off between number of targets and spectral coverage.

\textbf{Improved UV coatings and detectors:}
NASA is investigating new, enhanced LiF coatings
and improved, large-field microchannel plate detectors.
Both elements still lack flight qualification \citep{FFW17}.

\section{Conclusions}

Our partial understanding of extremely metal-poor massive stars jeopardizes
the interpretation of SNe and LGRBs,
star-forming galaxies throughout Cosmic History,
and the re-ionization epoch.
Teams around the world are working to provide a quantitative
characterization of these objects and realistic feedback prescriptions
that can be ingested by other Astrophysics disciplines.
Current efforts, focusing on nearby galaxies of the Local Group and vicinity,
are pushing current facilities beyond their limits.

In order to make sizable progress on this field
the wavelength coverage, sensitivity and spatial resolution of a 10m-class telescope in space is needed.
The LUVOIR observatory, one of four Decadal Survey Mission Concept Studies initiated in Jan 2016,
can potentially fulfill our technical requirements.
We propose that ESA joins NASA in the construction of LUVOIR,
building on the past and current synergies that continue making
HST an extraordinarily successful telescope.
Moreover, ESA can play a fundamental role in this quest by providing an optical spectrograph
that will be fundamental for LUVOIR's suite of instruments.

\pagebreak

\end{document}